# Improving Hydrodynamic Modeling of Free-Swimming Algae Using a Modified Three-Sphere Approach

Md Iftekhar Yousuf Emon[1], Gregorius R. Pradipta[2], Xiang Cheng[2,3], Xin Yong[1]*

[1]Department of Mechanical and Aerospace Engineering, University at Buffalo, Buffalo, NY 14260, USA

[2]Department of Chemical Engineering and Materials Science, University of Minnesota, Minneapolis, MN 55455, USA

[3]Saint Anthony Falls Laboratory, University of Minnesota, Minneapolis, MN 55414, USA

*E-mail: xinyong@buffalo.edu

**Abstract**

The beating flagella of the green alga *Chlamydomonas reinhardtii* play a prominent role in cellular mechanics, enabling cells to both displace and sense surrounding fluid. Specifically, flagellum-induced fluid transport enables microalgae to swim through fluid media and interact with other microorganisms. Minimal models, such as the three-sphere model with one sphere representing the cell body and two orbiting spheres mimicking the flagella, have been widely adopted to study various aspects of algal motility, including the synchronization of flagellar beating, run-and-tumble swimming, responses to shear flow, cell-body rolling, and helical navigation. However, detailed investigation of the algal flow fields generated by this minimal model remains limited. In this study, we systematically examine the time-averaged and time-resolved fluid flows generated by the three-sphere algae model and compare the numerical predictions with experimental data. Our findings reveal that the standard three-sphere model fails to produce key flow characteristics observed experimentally. To address this discrepancy, we explore a modified three-sphere model with refined flagellar beating dynamics and identify that differential drag acting on the flagellar spheres is the dominant factor influencing the fidelity of the modeled flow fields. These results advance the fundamental understanding of the flagellum–fluid interactions and algal flow and enhance our ability to accurately simulate microswimmer dynamics.

# 1. Introduction

Hydrodynamics at low Reynolds number (Re) is dominated by viscous forces, and fluid inertia becomes negligible (Happel J, 1965; Lauga & Powers, 2009; Purcell, 1977). In this regime, a swimmer that executes a time-reversible (reciprocal) stroke cannot achieve net displacement, a constraint famously encapsulated by Purcell's scallop theorem (Daddi-Moussa-Ider et al., 2023). Thus, microorganisms must generate non-reciprocal deformations via beating flagella, cilia, or coordinated body shape changes to break time-reversal symmetry and enable swimming at low Re (Shapere & Wilczek, 1987). Taylor's waving-sheet calculation first demonstrated locomotion without inertia in purely viscous flows (Taylor, 1951). Building on that insight, various minimal models have been developed that capture non-reciprocal strokes in low Re flow and the resulting hydrodynamic coupling through theoretical development (Dreyfus, Baudry, & Stone, 2005; Lighthill, 1952; Najafi & Golestanian, 2004) and controlled experiments (Dreyfus, Baudry, Roper, et al., 2005; Leoni et al., 2009; Tierno et al., 2008).

*Chlamydomonas reinhardtii* (*C. reinhardtii*), a unicellular green alga that swims with two identical anterior flagella in a breaststroke-like gait (Rüffer & Nultsch, 1985), has become a canonical model for studying microswimmer hydrodynamics (Cortese & Wan, 2021; Goldstein et al., 2009; Polin et al., 2009; Ruffer & Nultsch, 1998; Rüffer & Nultsch, 1985; Schaller et al., 1997; Tam & Hosoi, 2011; Wang & Tsang, 2025), including swimming kinematics and the resulting flow. Time-resolved and time-averaged flow measurements around biflagellate *C. reinhardtii* reveal a characteristic near-field structure: a front stagnation point, a pair of counter-rotating side vortices flanking the body, and phase-dependent reorganization of the flow around the flagella over the beat cycle. During the power stroke, the side vortices intensify and tilt rearward while the hyperbolic stagnation point shifts toward the body. During the recovery stroke, the flow weakens and partially reverses, reflecting the reversal of flagellar motion (Drescher et al., 2010; Guasto et al., 2010). Follow-up studies in confined geometries corroborated these near-field features and showed that geometric confinement substantially alters the spatiotemporal distribution of flagellar forcing, vortex structure, and mixing patterns, underscoring the sensitivity of near-field flows to stroke details and to the presence of nearby boundaries (Jeanneret et al., 2019; Mondal et al., 2021).

Inspired by *C. reinhardtii* swimming, a simple mechanical analogue, the three-sphere model, was introduced in two independent studies in 2012 (Bennett & Golestanian, 2013; Friedrich & Julicher, 2012). In this minimal model, one sphere represents the body and two spheres moving along prescribed orbits mimic the flagella, capturing the basic symmetry of *C. reinhardtii* and the coupling between internal actuation and hydrodynamic interactions that lead to net propulsion. This framework has been extended to study flagellar synchronization, three-dimensional (3D) helical swimming, and phototactic steering (Cortese & Wan, 2021; Polotzek & Friedrich, 2013). Prior works on three-sphere formulations of biflagellate swimmers also include the development of a bead-spring scaffold to study shear-flow responses and suspension rheology (Jibuti et al., 2017). Recent works adopted a *Chlamydomonas*-derived three-sphere swimmer with flagellar spheres transversing elliptical orbits and used it as a reduced-order platform to test locomotion strategies (Liu et al., 2025; Wang & Tsang, 2025).

Despite the widespread use of the three-sphere swimmer for swimming kinematics and flagellar synchronization, the flow generated by this minimal model remains largely unexplored. Because the flow fields mediate interactions with solid boundaries and neighboring swimmers, they are essential for assessing the fidelity of the model representing biflagellate swimmers. Here, we investigate time-resolved and time-averaged flow fields for the standard three-sphere swimmer and compare them directly with experimental measurements. In its original form, the model reproduces forward propulsion but misses key near-field flow signatures of breaststroke swimming observed in experiments. This gap motivates us to explore controlled extensions for improving the flow features while preserving the essential kinematics and simplicity of this minimal model.

First, inspired by non-circular flagellar paths observed in experiments, we generalize the flagellar trajectories from circles to ellipses and introduce mirror-symmetric in-plane rotation of elliptical axes relative to the swimming direction. We scrutinize how orbit shape and orientation affect swimming kinematics and the time-averaged flow. We also compare constant-torque actuation with phase-dependent torques, which have been used to model unequal power and recovery strokes in the studies of flagellar kinematics. Finally, motivated by the differential drag on slender flagella arising from shape variations during the power and recovery strokes, we implement a size-ratio modulation for the flagellar spheres that reduces effective drag during the recovery stroke and examines the alterations in flow fields and swimming kinetics. The modified model successfully captures the flow features that were previously missing while achieving higher swimming speed and efficiency. These results elucidate the nontrivial correlation between time-averaged flow and the non-reciprocity between power and recovery strokes, providing important insight into the accurate modeling of algal flow within a three-sphere framework.

## 2. Methodology

### 2.1 Model framework

The three-sphere swimmer is a simplified, theoretical model inspired by the biflagellate green algae *C. reinhardtii* (Friedrich & Julicher, 2012), developed to capture its fundamental symmetries and dynamics in a low Re environment as shown in figure 1a. This swimmer consists of three spherical beads interconnected by a frictionless, planar scaffold forming an isosceles triangle, characterized by its base length ($2l$) and height ($h$). Two identical spheres, representing the flagella, each of radius $r_a$ are attached to the base vertices of the scaffold, denoted by $S_1$ and $S_2$. The flagellar spheres move along circular trajectories of radius $R$ around $S_1$ and $S_2$, respectively parameterized by phase angles $\phi_1$ and $\phi_2$ prescribed by the right-hand rule. The third sphere, with radius $r_b$, is positioned at the apex of the triangle and rigidly attached to the scaffold, representing the cell body. In this study, we consider two-dimensional (2D) swimming where the entire system is confined to the $x$-$y$ plane. Thus, each sphere would have 3 degrees of freedom, two translational and one angular, giving a total of nine degrees of freedom. However, because the spheres are connected by a rigid scaffold, we describe the swimmer configuration using five generalized coordinates $\mathbf{q} = (x_0, y_0, \theta_0, \phi_1, \phi_2)^T$, where $x_0$ and $y_0$ give the center of the body sphere in the laboratory frame and $\theta_0$ is the right-handed rotation angle of the body sphere about the $z$-axis. The laboratory-frame positions of the three spheres are $\mathbf{r}_i = (x_i, y_i, 0)^T$, $i = 0, 1, 2$. To relate body-frame geometry to the laboratory frame, we introduce the planar rotation matrix

$$\mathbf{R}(\theta_0) = \begin{pmatrix} \cos\theta_0 & -\sin\theta_0 & 0 \\ \sin\theta_0 & \cos\theta_0 & 0 \\ 0 & 0 & 1 \end{pmatrix} \quad (1)$$

which rotates vectors from the body frame into the laboratory frame. In the body frame, the positions of the flagellar attachment points $S_1$ and $S_2$ are $\mathbf{s}_i^B = h\hat{\mathbf{e}}_1 + (-1)^i l \hat{\mathbf{e}}_2$, $i = 1, 2$, where $\hat{\mathbf{e}}_1 = (1,0,0)^T$ and $\hat{\mathbf{e}}_2 = (0,1,0)^T$ are the unit vectors in the body frame. The corresponding attachment points in the laboratory frame are thus $\mathbf{s}_i = \mathbf{r}_0 + \mathbf{R}(\theta_0) \cdot \mathbf{s}_i^B$, $i = 1, 2$. For circular-orbit simulations, the flagellar displacements in the body frame are

$$\mathbf{r}_i^B = R(\cos\phi_i\, \hat{\mathbf{e}}_1 + \sin\phi_i\, \hat{\mathbf{e}}_2), i = 1, 2 \quad (2)$$

so that the instantaneous laboratory-frame positions of the flagellar spheres are

$$\mathbf{r}_i = \mathbf{r}_0 + \mathbf{R}(\theta_0) \cdot (\mathbf{s}_i^B + \mathbf{r}_i^B), i = 1, 2 \quad (3)$$

We define the phase origin such that $\phi_i = 0$ when the $i$-th flagellar bead lies on the $+\hat{\mathbf{e}}_1$ direction in the body frame. For synchronized strokes, the phases $\phi_1 = -\phi_2$ with their magnitudes increasing monotonically from 0 to $2\pi$ over one beat cycle. Their time derivatives $\dot\phi_1$ and $\dot\phi_2$ define the phase speeds. Throughout this study, a positive phase velocity, $\dot\phi_i > 0$ corresponds to counterclockwise rotation of the $i$-th flagellum when viewed from the positive $z$-direction. In other words, $\dot\phi_1$ is negative and $\dot\phi_2$ is positive for the breaststroke gait (figure 1a).

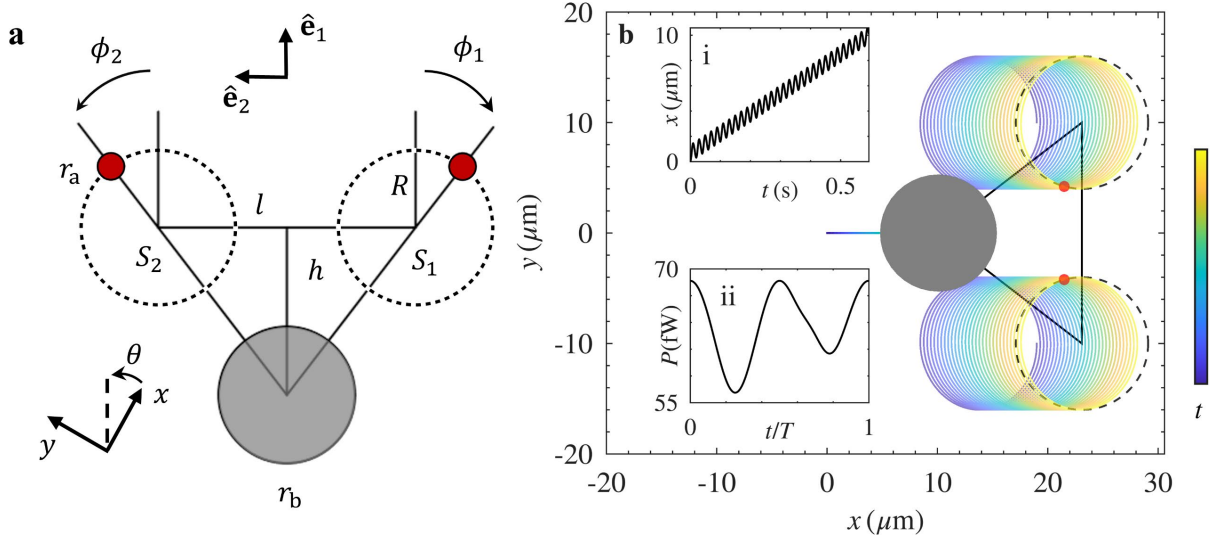

Figure 1: Three-sphere swimmer model and locomotion performance. (a) Schematic of the model geometry showing body sphere (grey) radius $r_b$, flagellar sphere (red) radius $r_a$, scaffold base length $2l$ ($S_1 - S_2$ distance), scaffold height $h$, and circular orbit radius $R$. The phase angles of the flagellar spheres, $\phi_1$ and $\phi_2$, are measured relative to $\hat{\mathbf{e}}_1$ in the body frame. (b) In-plane trajectories (lab frame) of the body and flagellar spheres over time. Insets are (i) instantaneous body displacement over time and (ii) instantaneous viscous dissipation over one beat cycle as a function of normalized time (with respect to beat period $T$).

Among the existing models, the Friedrich–Jülicher (F-J) variant (Friedrich & Julicher, 2012) formulates the dynamics with constant driving torques applied to the flagellar spheres, emphasizing synchronization via hydrodynamic friction forces. In contrast, the Bennett–Golestanian variant (Bennett & Golestanian, 2013) incorporates a stroke-dependent forcing profile together with intrinsic noise, enabling nonlinear dynamics that reproduce the experimentally observed stochastic run-and-tumble gait. Here, we adopt the F-J formulation because it uses a symmetric generalized friction matrix with separate hydrodynamic and internal contributions. Internal dissipation in the flagellar driving mechanism is added as a simple diagonal contribution acting only on the phase variables, so hydrodynamic dissipative and internal energy losses are cleanly separated. This formulation makes it straightforward to incorporate geometric and driving modifications by updating the constraint matrix and the mobility tensor (described in Section 2.3), without changing the overall structure of the equations of motion. Compared to stroke-prescribed formulations, the F-J approach keeps the coupling between body motion, flagellar phase dynamics, and hydrodynamic interactions explicit and symmetric, which is advantageous for analyzing both flow features and kinematics. Below, we briefly describe the hydrodynamic interactions and dynamics of the F-J formulation and additional details can be found in (Polotzek & Friedrich, 2013).

## 2.2 Hydrodynamic interactions

The swimmer operates in the Stokes flow regime (Re $= \rho v l / \mu \ll 1$) where inertial forces are negligible compared to viscous forces (Happel J, 1965; Lauga & Powers, 2009). Here, $l$ is the length scale of the swimmer, $v$ is the average swimming speed, $\rho$ is the density of fluid and $\mu$ is its viscosity. For *C. reinhardtii*, $l$ is about 10 μm and the average swimming velocity is approximately 100 μm/s and therefore Re ~ $10^{-3}$. In this regime, hydrodynamics is governed by the Stokes equation, and the translational and rotational velocities of spherical beads depend linearly on the hydrodynamic forces and torques acting on them.

The translational and angular velocities of the three spheres are grouped into a 9-component vector

$$\dot{\mathbf{q}}_0 = (u_0, v_0, \omega_0, u_1, v_1, \omega_1, u_2, v_2, \omega_2)^T \quad (5)$$

and the corresponding forces and torques are

$$\mathbf{P}_0 = (F_{0x}, F_{0y}, T_0, F_{1x}, F_{1y}, T_1, F_{2x}, F_{2y}, T_2)^T \tag{6}$$

These are related by the grand mobility and friction matrices

$$\dot{\mathbf{q}}_0 = \mathbf{M}_0 \cdot \mathbf{P}_0 \tag{7}$$

where $\mathbf{M}_0 \in \mathbb{R}^{9\times 9}$ is the grand mobility matrix for the three spheres and $\mathbf{\Gamma}_0 = \mathbf{M}_0^{-1}$ is the corresponding grand friction matrix. The diagonal blocks of $\mathbf{\Gamma}_0$ represent the Stokes drag and rotational resistance of isolated spheres, while the off-diagonal blocks encode hydrodynamic couplings between distinct spheres. To account for different sizes of flagellar and body spheres, the entries of $\mathbf{M}_0$ are constructed using the pairwise Rotne–Prager–Yamakawa (RPY) tensor, a pairwise approximation to the Stokes mobility for finite-sized spheres (Zuk et al., 2014). The RPY tensor regularizes the flow near each bead, incorporates leading finite-radius corrections for translation–translation, translation–rotation, and rotation–rotation couplings, and guarantees a symmetric positive-definite mobility (and hence friction) matrix even when the spheres are not widely separated.

The instantaneous total dissipation rate of this dynamic system is

$$\mathcal{R}_h = \dot{\mathbf{q}}_0^T \cdot \mathbf{\Gamma}_0 \cdot \dot{\mathbf{q}}_0 \tag{8}$$

This dissipation counts for all three spheres and sphere-sphere interactions and is non-negative for all admissible motions, so $\mathbf{\Gamma}_0$ is symmetric positive definite (or positive semidefinite in the presence of exact constraints). This expression of hydrodynamic dissipation readily allows one to calculate the swimming efficiency as detailed in Section 2.3.

## 2.3 Swimming kinematics

The swimmer configuration is more conveniently described by the five generalized coordinates $\mathbf{q} = (x_0, y_0, \theta_0, \phi_1, \phi_2)^T$ introduced in Section 2.1. A Jacobian matrix $\mathbf{L}$ with entries $L_{ij} = \partial(q_0)_i / \partial(q)_j$ is introduced to which maps generalized velocities and force $\dot{\mathbf{q}}$ and $\mathbf{P}_0$ to the velocities and forces/torques of individual spheres, $\dot{\mathbf{q}}_0 = \mathbf{L} \cdot \dot{\mathbf{q}}$ and $\mathbf{P}_h = \mathbf{L}^T \cdot \mathbf{P}_0$. Substituting these relations into hydrodynamic dissipation in terms of generalized coordinates,

$$\mathcal{R}_h = \dot{\mathbf{q}}^T \cdot \mathbf{\Gamma}_h \cdot \dot{\mathbf{q}} \tag{9}$$

Through this transformation, we define a reduced hydrodynamic friction matrix $\mathbf{\Gamma}_h \in \mathbb{R}^{5\times 5}$, given by

$$\mathbf{\Gamma}_h = \mathbf{L}^T \cdot \mathbf{M}_0^{-1} \cdot \mathbf{L} \tag{10}$$

which can be calculated according to the instantaneous positions of the three spheres. To represent energy losses in the internal flagellar driving mechanism, a separate internal friction contribution acting only on the phase angle is introduced, $\mathbf{\Gamma}_\kappa = \text{diag}(0,0,0,\kappa,\kappa)$, where $\kappa$ is a dimensionless phase friction parameter. The total friction matrix and the equation of motion with respect to the generalized coordinate are $\mathbf{P} = (\mathbf{\Gamma}_h + \mathbf{\Gamma}_\kappa) \cdot \dot{\mathbf{q}}$. Given that the swimmer is force- and torque-free, we have $\mathbf{P} = (0,0,0,m_1,m_2)^T$ with $m_i$ being the active torques applied on the flagellar spheres. The final equation of motion reads

$$(0,0,0,m_1,m_2)^T = (\mathbf{\Gamma}_h + \mathbf{\Gamma}_\kappa) \cdot \dot{\mathbf{q}} \tag{11}$$

Integrating this equation of motion yields the full trajectories of the swimmer.

The swimming efficiency is quantified using the Lighthill hydrodynamic efficiency, defined as the ratio of the useful towing power required to drag the body sphere at the swimmer's mean swimming speed to the average viscous dissipation generated by the swimmer (Lauga, 2020; Lighthill, 1952; Nasouri et al., 2019). The useful towing power is $P_{\text{drag}} = 6\pi\mu r_b \langle U \rangle^2$, where $\langle U \rangle = (1/T) \int_0^T U(t)dt$ with $T$ being the period of a beat cycle. The instantaneous viscous dissipation is computed from the generalized hydrodynamic resistance matrix as $P(t) = \dot{\mathbf{q}}_f^T \mathbf{P}_f$ with $\dot{\mathbf{q}}_f = (u_1, v_1, \omega_1, u_2, v_2, \omega_2)^T$ and $\mathbf{P}_f = (F_{1x}, F_{1y}, T_1, F_{2x}, F_{2y}, T_2)^T$, taking only the active contributions from two flagellar spheres. The swimming efficiency $\eta$ therefore read

$$\eta = \frac{6\pi\mu r_b \langle U \rangle^2}{\langle P \rangle} \tag{12}$$

Here, $\langle P \rangle = (1/T) \int_0^T P(t)dt$ is the cycle-averaged dissipation.

## 2.4 Flow field calculation

The instantaneous fluid velocity field at each point in space is calculated by superposing the contributions from the body sphere and the two flagellar spheres. The flow induced by one sphere is written using fundamental singularities in the Stokes equations (Polotzek & Friedrich, 2013), as the superposition of a Stokeslets, a potential dipole, and a rotlet, which captures the effects of both translational and rotational motion of the sphere. Specifically, the velocity at a field point $\mathbf{r}$ due to all spheres is

$$\mathbf{v}(\mathbf{r}) = \sum_{i=0}^{2} \left[ \frac{3a_i}{4d_i} (\mathbf{I} + \mathbf{e}_i \mathbf{e}_i^T) \cdot \mathbf{U}_i + \frac{a_i^3}{4d_i^3} (\mathbf{I} - 3\mathbf{e}_i \mathbf{e}_i^T) \cdot \mathbf{U}_i + \frac{a_i^3}{d_i^2} (\mathbf{\Omega}_i \times \mathbf{e}_i) \right] \tag{13}$$

Here, $a_i$ is the radius of the sphere, $d_i = |\mathbf{r} - \mathbf{r}_i|$ is the distance between the field point and the sphere center, $\mathbf{e}_i = (\mathbf{r} - \mathbf{r}_i)/d_i$ is the unit vector, $\mathbf{U}_i$ the translational velocity, and $\mathbf{\Omega}_i$ is the angular velocity of the sphere.

## 2.5 Parameterization and nondimensionalization

The parameters used in our simulation are adopted from a recent study by Cortese and Wan (Cortese & Wan, 2021), who refined the three-sphere model using experimentally measured geometric and dynamical features of wild-type *C. reinhardtii*. All lengths are nondimensionalized by the characteristic length scale $l_0$. The sphere representing the body has a radius of 0.53 and the two flagella spheres have a radius of 0.05. The scaffold triangle has a base length of 2 and a height of 1.3. Each flagellar bead rotates along a circular orbit of radius 0.6. To report physical values, we map simulations to experiments using the physical length scale $l_0 = 10$ μm and the typical 50 Hz beat frequency of *C. reinhardtii* (Li et al., 2025; Merchant et al., 2007). The non-dimensional beat frequency, $f = 1/T$, is obtained by averaging over the last ten cycles, identified using zero crossings of $U$. This frequency is then used to extract the model time scale, $\tau = 50/f$ s. With this mapping, dimensional velocities are obtained as

$$U_{\text{phys}} = \frac{Ul}{\tau} \tag{14}$$

## 2.6 Orbit shape modification

Motivated by experimental observations of non-circular flagellar trajectories (Bayly et al., 2010; Kurtuldu et al., 2013; Liu et al., 2025), we explore the effect of stroke geometry. We generalize these orbits to ellipses with major semi-axis $R_a$, minor semi-axis $R_b$, aspect ratio $\gamma$, and in-plane orientation angle $\alpha$. We parameterize the orbit by a phase angle $\phi_i(t)$ and define an effective radius

$$\tilde{R}(\phi_i) = \left[\frac{\cos^2 \phi_i}{R_a^2} + \frac{\sin^2 \phi_i}{R_b^2}\right]^{-\frac{1}{2}} \tag{15}$$

The orbit eccentricity is controlled by the aspect ratio, $e = \sqrt{1 - \gamma^2}$, where $\gamma = R_b/R_a$. The in-plane orientation of each ellipse's major axis about the $z$-axis is specified by $\alpha$, following the same right-hand rule as other rotation angles. To keep the orbits mirror-symmetric, we introduce rotated basis vectors for the orbit of flagellar sphere 1,

$$\hat{\mathbf{e}}_1^1 = (\cos \alpha, \sin \alpha, 0)^T, \hat{\mathbf{e}}_2^1 = (-\sin \alpha, \cos \alpha, 0)^T \tag{16}$$

and for flagellar sphere 2,

$$\hat{\mathbf{e}}_1^2 = [\cos(-\alpha), \sin(-\alpha), 0]^T, \hat{\mathbf{e}}_2^2 = [-\sin(-\alpha), \cos(-\alpha), 0]^T \tag{17}$$

The flagellar displacements in the body frame along the elliptical orbits are thus changed to

$$\mathbf{r}_i^B = \tilde{R}(\phi_i)\big(\cos \phi_i \, \hat{\mathbf{e}}_1^i + \sin \phi_i \, \hat{\mathbf{e}}_2^i\big), i = 1, 2 \tag{18}$$

## 2.7 Active driving and phase-dependent torques

Our implementation of the three-sphere model based on the framework introduced by Friedrich and Jülicher (Friedrich & Julicher, 2012) assumes a constant driving torque throughout the flagellar cycle. In contrast, several other variants of the three-sphere model have incorporated phase-dependent torques or forces, presumably aiming to more accurately account for the temporal asymmetries observed in real flagellar motion (Bennett & Golestanian, 2013; Liu et al., 2025).

To evaluate the effect of phase-dependent torques, we prescribe

$$m_i(\phi_i) = m_i^0[1 + c_i \cos(\phi_i + \psi_i)], i = 1, 2 \quad (19)$$

where $m_1^0 = -m_2^0$ are opposite mean torques on the two flagella. Here, $c_1$ and $c_2$ are modulation amplitudes that control the strength of phase variation, and $\psi_1$ and $\psi_2$ are phase offsets. We examined two phase-dependent torque models consistent with those reported in previous studies for the circular orbit (Bennett & Golestanian, 2013; Liu et al., 2025; Wang & Tsang, 2025). In Model A, the stroke with stronger force/torque corresponds to the upper half of the orbit with $\psi_1 = -\frac{\pi}{2}$ and $\psi_2 = \frac{\pi}{2}$, whereas in Model B, it corresponds to the lower half with $\psi_1 = \frac{\pi}{2}$ and $\psi_2 = -\frac{\pi}{2}$. In both models, $c_1 = c_2 = 0.7$.

**2.8 Flagellar sphere size modification**

We introduce an additional modification to the model by varying the size of the flagellar spheres during the stroke cycle. Specifically, we aim to enhance the flow asymmetry between the power and recovery strokes, as required to reproduce experimentally observed flow features. One effective way to accomplish this is to reduce hydrodynamic drag during the recovery stroke. During the power stroke, the flagella are extended and oriented predominantly perpendicular to the swimming axis, whereas during the recovery stroke, they fold closer to the body and retract with a significantly reduced lateral footprint (figure S1a). This behavior is reminiscent of the hydrodynamics of slender filaments, which experience significantly higher drag when oriented perpendicular to the flow than when aligned parallel (Lauga, 2020).

To capture this anisotropic drag in a minimal way, the effective radius of the flagellar spheres becomes dependent on the phase, while the body radius $r_b$ is fixed. We define a reference flagellar radius $r_a^0$ and a dimensionless recovery-to-power size ratio $\lambda$ and prescribe

$$r_a(\phi_i) = \begin{cases} r_a^0 & 0 \le |\phi_i| < \pi \\ \lambda r_a^0 & \pi \le |\phi_i| < 2\pi \end{cases} \quad (20)$$

with periodic extension to all $\phi_i$. The interval $0 \le |\phi_i| < \pi$ corresponds to the power stroke and $\pi \le |\phi_i| < 2\pi$ to the recovery stroke. The hydrodynamic mobility tensor is updated at each time step using the current value of $r_a(\phi_i)$, so that both the self-mobility and the pairwise RPY interactions between the spheres vary over the beat cycle in response to the varying size of the flagellar spheres.

**2.9 Scaffold geometry modification**

In the standard three-sphere model, the flagellar spheres are attached by a scaffold with a vertical offset $h$ in front of the body and lateral offset $l$, therefore always creating the vortices ahead of the body. But the previous experimental results show that the vortices form by the side of the body (Drescher et al., 2010; Guasto et al., 2010; Pradipta et al., 2025). To test whether the experimentally observed side vortex position can be recovered, we systematically vary the scaffold height $h$ while keeping the flagellar "root length" fixed, i.e., maintaining $\sqrt{l^2 + h^2}$ constant by adjusting $l$ accordingly. We combine scaffold geometry modification with elliptical orbits and flagellar sphere size variation to optimize the time-averaged flow field against experiments. These geometric modifications do not alter the governing Stokes hydrodynamics or the resistance-based formulation; they only change the kinematic mapping between generalized coordinates and bead velocities through the orbit parametrization.

### 2.10 Exploratory two-swimmer simulations

To assess whether the proposed modifications influence alga–alga interactions, we perform preliminary leader–follower simulations of two identical swimmers in an unbounded 3D flow. Each swimmer in the two-swimmer system obeys the same generalized resistance formulation used in the single-swimmer model, with hydrodynamic couplings between all six spheres included through the full grand mobility matrix. Swimmers are initialized with the same phase and separated by an initial streamwise offset $\Delta x_0$. The primary observable is the temporal evolution of body–body separation distance $\Delta x(t)$.

## 3. Results

### 3.1 Flow of the standard three-sphere model

We first validate the sphere trajectory and displacement of the breaststroke swimmer using the standard three-sphere model (figure 1b). The swimmer moves in a straight line with an oscillatory forward-backward motion. The net displacement of this swimming mode is forward, as shown in figure 1b(i), with an average velocity of 15.72 µm/s. This value is substantially lower by roughly an order of magnitude than experimentally reported swimming speeds of *C. reinhardtii* ~100 µm/s (Amador et al., 2020; Drescher et al., 2010). The swimmer's instantaneous viscous dissipation (figure 1b(ii)) shows that the average dissipation is slightly higher in the recovery stroke than in the power stroke, which is also significantly different from the experimental results (Guasto et al., 2010; Pradipta et al., 2025).

To assess the model's fidelity in predicting not only the swimming kinematics but also the flow field around an alga, we evaluate the time-resolved flow fields (movie 1) at six corresponding instances within the beat cycle (figure 2), matching the phase locations reported by Guasto et al. (Guasto et al., 2010) to enable a direct comparison between the model and their experimental results. The first five snapshots capture various stages of the power stroke, while the final snapshot corresponds to the recovery stroke. The flow fields show agreement with experimental observations in reproducing the front stagnation point along the swimming axis and the rear-tilted vortices flanking the swimmer body during the initial phases of the power stroke (figure 2b). However, significant deviations can be observed past the peak of the power stroke. The side vortex structures rapidly diminish in the later phases of the power stroke (figure 2c-e). A rear stagnation point appears much earlier than in the experiments (figure 2c-d). Additionally, the experimentally observed positive force dipole (pusher) flow (see figure 3f in Ref. (Guasto et al., 2010)) and the front stagnation point are absent during the recovery phase (figure 2f). The lower branch of the velocity–phase profile (the left inset of figure 2) is less smooth, reflecting stronger hydrodynamic interactions when the spheres approach each other during the recovery stroke. The area enclosed by this lower branch, which corresponds to backward motion, is larger in the model than in the experiments, leading to a smaller net forward displacement over a cycle.

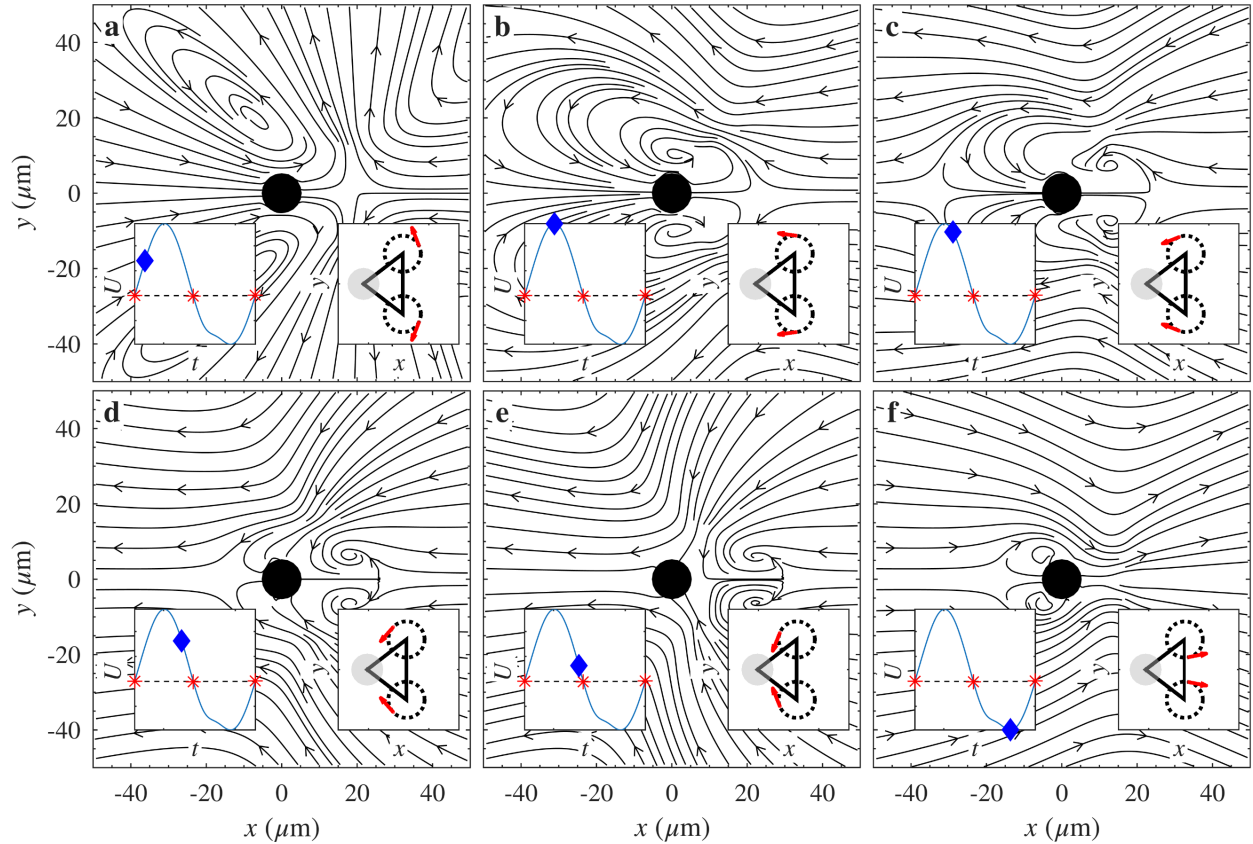

Figure 2: Time-resolved flow fields of the three-sphere swimmer over one beat cycle. Each panel (a–f) represents a distinct phase, showing the temporal evolution of the flow. The bottom-right inset in each panel illustrates the instantaneous morphology of the swimmer, i.e., the relative positions of the three spheres, where the red arrows indicate the translational velocity direction of the flagellar spheres. The bottom-left inset plots signed cell velocity (positive values correspond to forward motion) over a beat cycle and the specific phase (blue diamond), allowing direct comparison between swimmer kinematics and flow structure. The dashed line indicates $u = 0$ and the red stars mark the start and end of the power/recovery strokes.

Subsequently, we compute the time-averaged flow field over a full beat cycle (figure 3). At the orbit positions of flagellar spheres, two side vortices are present and oriented almost perpendicular to the axis of the swimming direction. A hyperbolic stagnation point appears at the tip of the side vortex. There is no sign of the front stagnation point. In contrast, the experimental measurements (Drescher et al., 2010; Guasto et al., 2010; Pradipta et al., 2025) registered a clear stagnation point in the front and side vortices tilted rearward with no attached stagnation points in the beat-cycle averaged flow. The cycle-average flow closely resembles the instantaneous flow early in the power stroke, indicating the dominance of the power stroke flow in the experiments. Both key flow features are absent in the flow field generated by the standard model. These discrepancies highlight the limitations of the standard three-sphere model and motivate modifications that can more faithfully reproduce the experimentally observed flow-field structure.

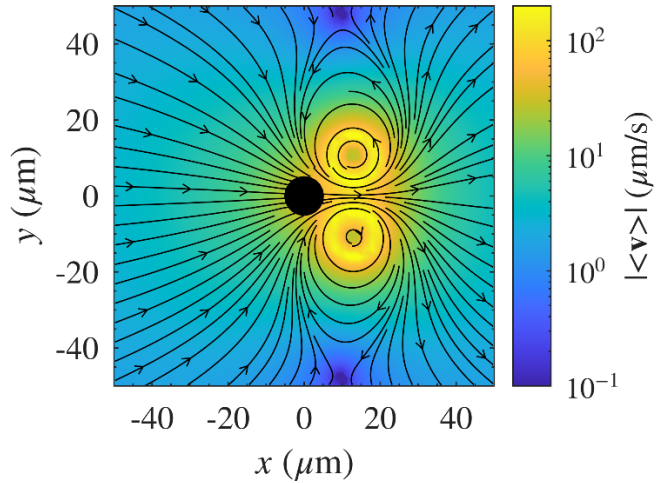

Figure 3: Time-averaged flow field over one beat cycle for the standard three-sphere model. The colormap represents the magnitude of the time-averaged flow.

## 3.2 Effects of orbit shape

Inspired by experimental observations of flagellar centroid trajectories (figure S1a) and non-circular orbits in other numerical models (Jibuti et al., 2017; Li et al., 2025), we investigate the influence of orbit shape on the swimmer's hydrodynamic performance and flow field structure. We examine the time-averaged flow fields, orbit shapes, and associated vortex cores for four different combinations of orbit parameters: two aspect ratios (0.3 and 0.8) and two orientation angles (30° and -30°), as shown in figure 4. The flow field is largely unchanged in the far field, outside the vortex cores. The vortices still remain approximately perpendicular to the swimming axis for all cases. In the near field, the effect of orbit shape is appreciable. To assess the influence of orbit shape on the vortex core, we conduct an analysis based on the local velocity gradient tensor,

$$\nabla \mathbf{v} = \begin{bmatrix} \frac{\partial v_x}{\partial x} & \frac{\partial v_y}{\partial x} \\ \frac{\partial v_x}{\partial y} & \frac{\partial v_y}{\partial y} \end{bmatrix} \quad (21)$$

This tensor was decomposed into a symmetric rate-of-strain tensor $\mathbf{S} = [\nabla \mathbf{v} + (\nabla \mathbf{v})^T]/2$ and an antisymmetric vorticity tensor $\mathbf{\Omega} = [\nabla \mathbf{v} - (\nabla \mathbf{v})^T]/2$. These components were used to construct a symmetric tensor, $\mathbf{T} = \mathbf{S}^2 + \mathbf{\Omega}^2$, proposed as a robust, Galilean-invariant measure of local flow topology (Hunt et al., 1988). The eigenvalues of $\mathbf{T}$, denoted $\lambda_1 \geq \lambda_2$, quantify the local rate of deformation and rotation. In this study, we focus on the dominant eigenvalue field $\lambda_1$ to delineate coherent vortex structures. Vortex cores are visualized by extracting the zero-contour of $\lambda_1$, which marks the boundary between regions of intense shear and rotation and has been shown to correlate well with the center of vortical structures (Chakraborty et al., 2005; Zhou et al., 1999). Our results reveal that orbit geometry significantly alters the shape of the vortex cores. However, beyond the core region, the overall structure of the flow field remains unchanged. Thus, this result suggests that modulating orbit shape alone does not improve the flow field.

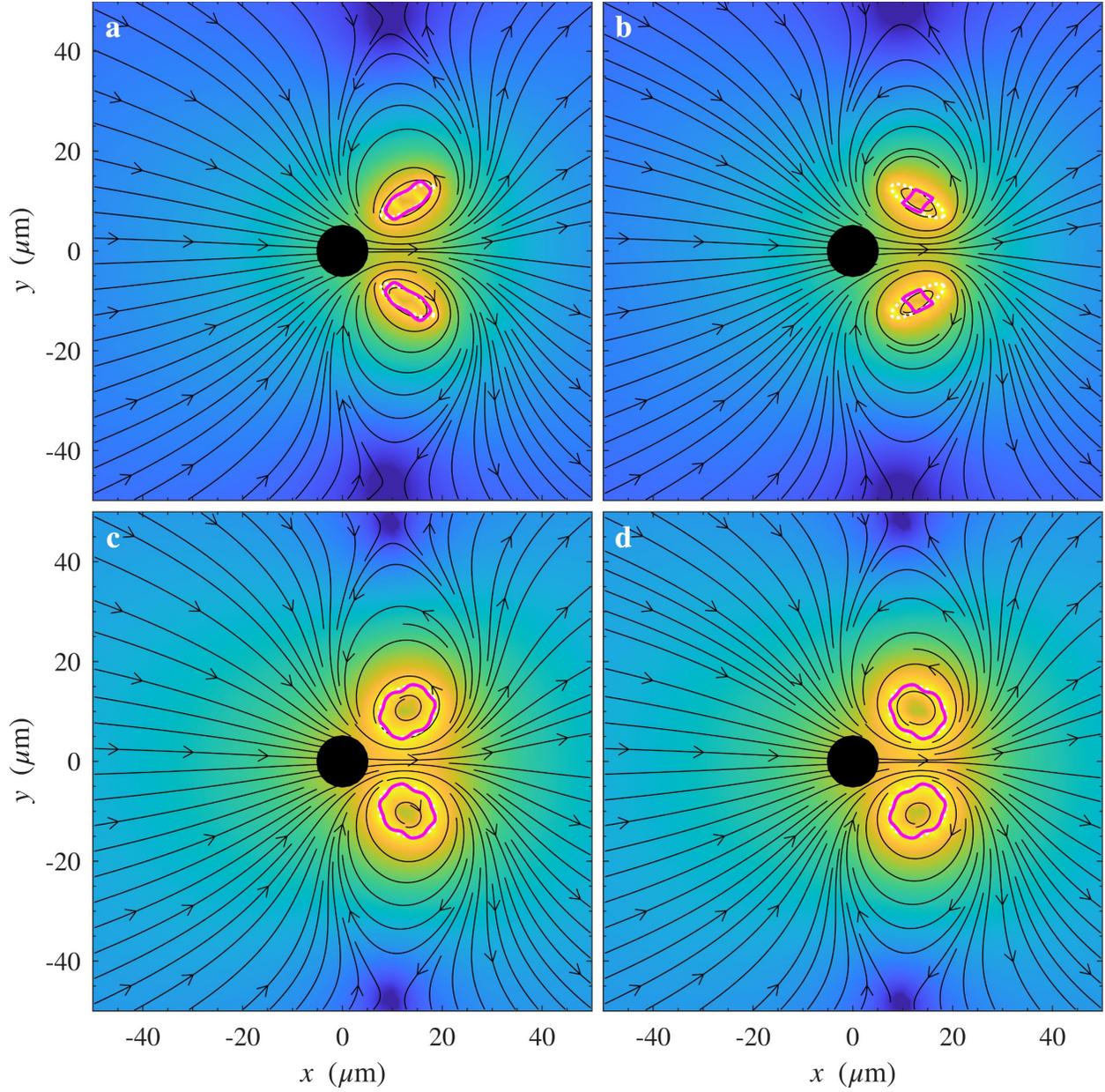

Figure 4: Time-averaged flow fields showing the effect of orbit orientation ($\alpha$) and aspect ratio ($\gamma$). Flagellar orbits are drawn in white dotted lines and the vortex core boundaries are highlighted in magenta. The top row (a, b) corresponds to $\gamma = 0.3$, and the bottom row (c, d) to $\gamma = 0.8$. The left column (a, c) shows results for $\alpha = -30°$, while the right column (b, d) corresponds to $\alpha = +30°$. The colormap is the same as in figure 3.

To understand the impact of orbit shape on propulsion, we also calculate the swimming speed and efficiency across a broad parameter space. The aspect ratio and orientation angle were respectively varied from 0.1 to 0.9 and -90° to 90°, covering a wide range of elliptic configurations.

Figure 5 shows that both swimming speed and efficiency exhibit only modest sensitivity (up to 19% and 48% variations, respectively) to orbit orientation and the model achieves its highest efficiency and swimming speed as the aspect ratio approaches one, corresponding to a circular trajectory.

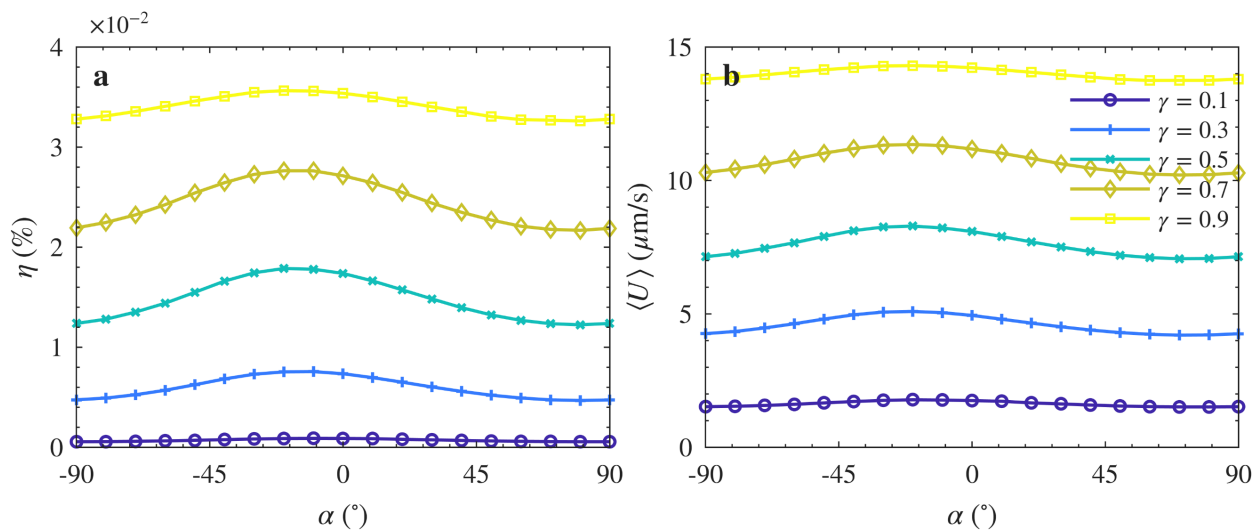

Figure 5: Variations in (a) swimming efficiency and (b) mean swimming velocity of the three-sphere model as a function of orbit orientation angle, for different orbit aspect ratios.

### 3.3 Constant vs phase-dependent torques

We then compare the swimming performance of two phase-dependent models (Section 2.7) with the constant torque reference. The swimming speeds are found to be comparable (table 1). This result demonstrates that different torque models do not significantly change velocity in physical units. Therefore, our findings disagree with one of the conclusions made by the previous work, which reported that the maximum average swimming speed is achieved under a constant driving force profile (Bennett & Golestanian, 2013). We note that the dimensionless velocity is significantly higher with constant torque, but the beating frequency also increases, resulting in a similar physical velocity. However, our results confirm that the efficiency of the constant-torque swimmer is approximately 37.6% and 41.6% higher than that of the respective phase-dependent model.

| Model | Mean velocity (model units) | Frequency (model units) | Physical velocity (μm/s) | Efficiency (%) |
|---|---|---|---|---|
| Standard | 0.0037 | 0.1188 | 15.7225 | 0.0388 |
| Model A | 0.0027 | 0.0848 | 15.7555 | 0.0282 |
| Model B | 0.0027 | 0.0853 | 15.6905 | 0.0274 |

Table 1: Comparison of mean forward velocity (model units), beat frequency (model units), and resulting physical swimming velocity (μm/s) for the standard model and two modified models (Model A and Model B). The standard model exhibits a higher mean velocity, higher beat frequency compared to the modified models and higher efficiency, while all three conditions produce nearly identical physical swimming speeds.

### 3.4 Effects of flagellar sphere size

Experimental measurements demonstrate a pronounced asymmetry between the power and recovery strokes. Over beat cycles, the flagellar configuration evolves along a conserved, phase-dependent loop: the flagella are extended with a large lateral excursion during the power stroke, whereas in the recovery stroke, they fold toward the cell body and retract with a substantially reduced lateral footprint (figure S1a). The magnitudes of forces reconstructed at the corresponding phases using the resistive force theory are substantially larger during the power stroke ($t/T < 0.5$) than during the recovery stroke ($t/T \geq 0.5$) (figure S1b). This pronounced power–recovery asymmetry in both flagellar kinematics and forcing provides direct quantitative evidence that the flagellar beat is intrinsically non-reciprocal. To fundamentally improve agreement between simulated and experimental flow fields, we introduce another feature to the model that treats the flagellar sphere size as a phase-dependent variable (Section 2.8). The time-averaged flow field with this modification is shown in figure 6. Unlike the standard model, the modified version successfully reproduces key features consistent with experimental observations in *C. reinhardtii* swimming (Drescher et al., 2010; Pradipta et al., 2025). The front stagnation point is present in the field of view, and the side vortices are swept toward the posterior. To quantify these features, we locate the front stagnation point and extract the vortex orientation angle. The latter is estimated by tracing a flow streamline that passes outside the core of the vortex. The orientation vector is then computed as the direction between its initial and final points. For consistency, the same initial streamline point is used across all size ratio cases.

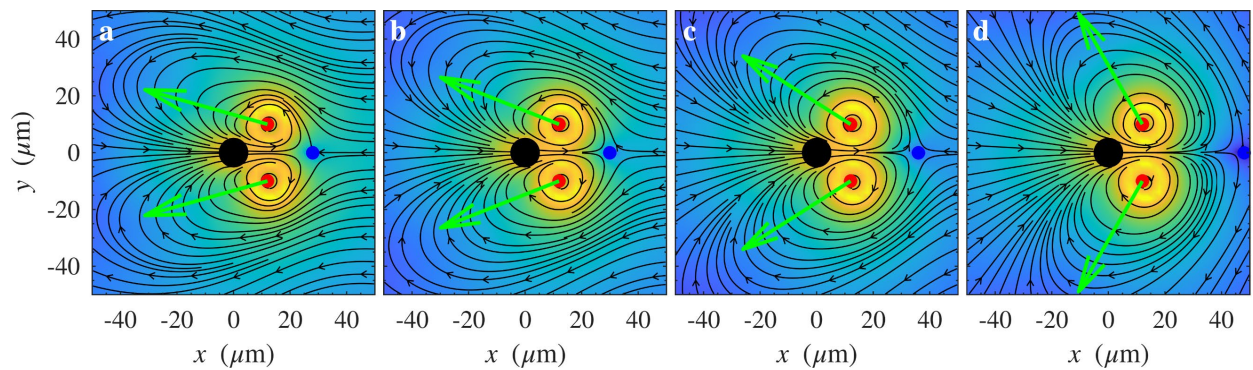

Figure 6: Time-averaged flow fields showing the influence of flagellar sphere size ratio on vortex orientation and stagnation point position for the value of the size ratio, $\lambda$ as (a) 0.3 (b) 0.4 (c) 0.5 and (d) 0.6 across panels. Streamlines depict flow direction, while the colormap indicates flow magnitude on a logarithmic scale, as in figures 3 and 4. Vortex orientations are highlighted with green arrows, and stagnation points are marked with blue dots.

To systematically assess the effect of the size ratio on flow structure, we vary it from 0.25 to 0.6 (figure 7a). Ratios above 0.6 are excluded as the stagnation point moves outside the field of view, while ratios below 0.25 are avoided due to the appearance of the rear stagnation point, which is not observed in the experiments (Drescher et al., 2010; Pradipta et al., 2025). The results reveal that decreasing the size ratio causes the front stagnation point to move closer to the swimmer body, while the orientation angle of the vortices increases relative to the swimming direction. We also calculate the swimming speed and efficiency for each size ratio (figure 7b). Both metrics decrease monotonically with increasing size ratio.

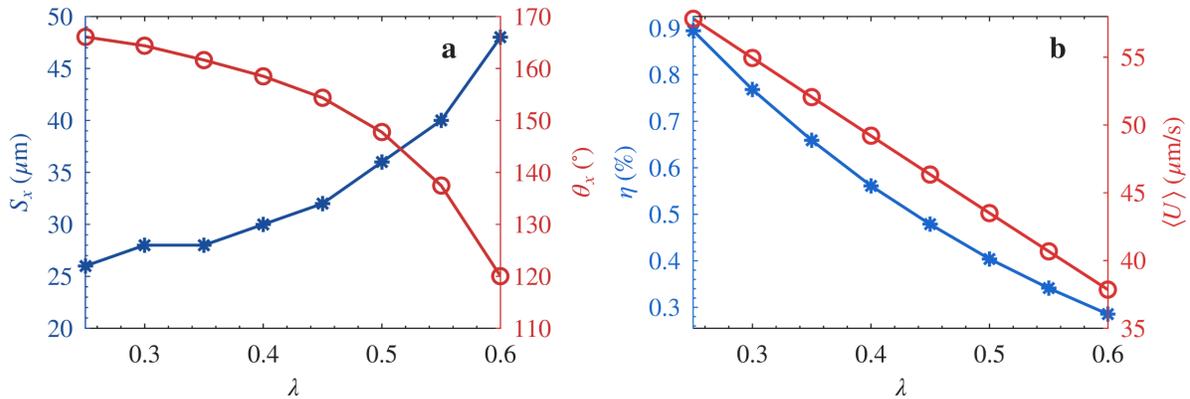

Figure 7: Quantitative influence of flagellar sphere size ratio, $\lambda$, on flow features and swimming kinematics. (a) Variations in stagnation point position $S_x$ and vortex orientation angle $\theta_x$ with size ratio. (b) Dependence of swimming efficiency $\eta$ and mean swimming velocity $\langle U \rangle$ on size ratio.

### 3.5 Scaffold and orbit modifications for side vortex features

Experiments consistently report vortex cores displaced laterally from the body rather than located ahead of it, motivating targeted geometric modifications of the scaffold and orbit. Reducing the scaffold height shifts the vortices but does not, by itself even with differential drag ($\lambda < 1$), reproduce the experimentally observed side vortex arrangement. Specifically, the dominant recirculation remains compressed and aligned with the swimming axis across all tested size ratios (see figure S3). However, upon modifying the scaffold with $h = 0.4$ and introducing a rotated elliptical orbit ($\gamma = 0.5$ and $\alpha = 45°$) along with size ratio $\lambda = 0.3$, the model produces a tilted side vortex pair and a stagnation point position highly consistent with the experimental flow (figure 8). We also find swimming velocity 46 μm/s and efficiency 0.87% for this optimized case.

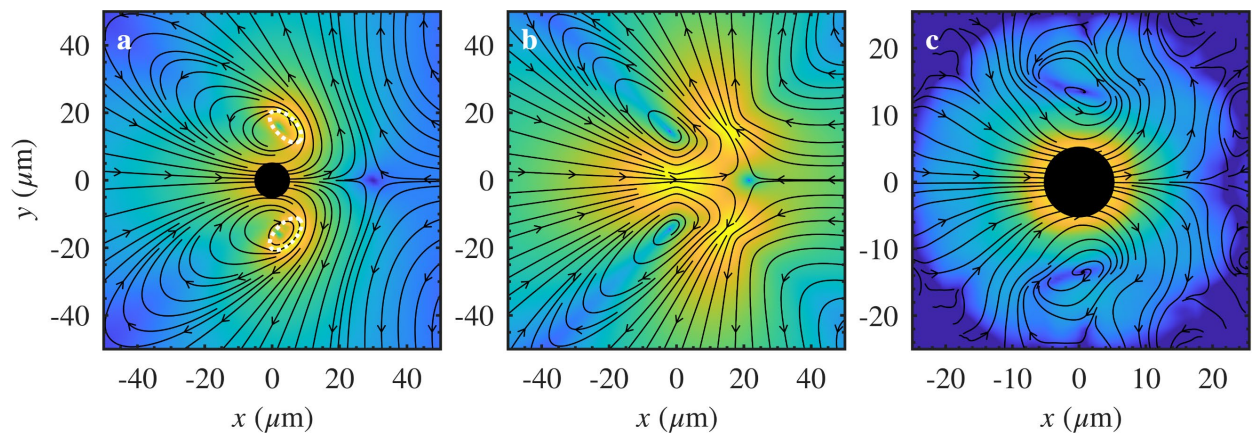

Figure 8: Time-averaged flow field over one flagellar beat cycle of (a) Modified three-sphere model with optimized scaffold geometry, orbit trajectory, and phase-dependent flagellar size ratio. The white dotted line indicates the flagellar orbit. The scaffold height is $h = 0.4$ (with $\sqrt{l^2 + h^2}$ held constant), producing side-oriented flagellar orbits. The orbit aspect ratio and orientation are $\gamma = 0.5$ and $\alpha = 45°$, respectively. The recovery-to-power size ratio is $\lambda = 0.3$. (b) Flow field produced by a static three-Stokeslet approximation consisting of two flagellar forces and a body force satisfying the force-free condition. The body force is located at $(0,0)$ and points along $\hat{\mathbf{e}}_1$, while the two flagellar forces are positioned at $(15,15)$ and $(15,-15)$ μm and oriented outward at $\pm 135°$ relative to $\hat{\mathbf{e}}_1$. Flagellar force magnitude corresponds to the mean value of the full cycle in figure S1b. (c) Experimental cycle-averaged flow field reconstructed from 3D imaging of freely swimming *C. reinhardtii.*, obtained by particle-tracking velocimetry of 1~μm tracer particles; velocity vectors are binned into $1 \times 1 \times 1$ voxels and averaged over one beat cycle.

## 4. Discussion and Conclusion

Our investigation into the standard three-sphere model and its successive modifications offers new insights into the hydrodynamic mechanisms governing microswimmer flow fields. The displacement-time plot for the reference model (figure 1b) reveals a key feature of non-reciprocal motion: the swimmer advances more during the power stroke than it retreats during the recovery stroke, thereby breaking time-reversal symmetry. Because the model drastically simplifies a slender flagellum as a single sphere, this asymmetry arises only from non-reciprocal hydrodynamic coupling between the spheres. The coupling is stronger during the recovery stroke, when the flagellar and body spheres are in closer proximity. Consistent with this coupling argument, the phases that exhibit stronger near-field interactions also demand larger power input to execute the stroke. Accordingly, the standard model predicts modestly higher viscous dissipation during the recovery stroke than during the power stroke, contrary to experimental trends.

The standard model reproduces certain key experimental features in the phase-resolved flow field during the initial phase of the power stroke, notably the formation of front stagnation points and counter-rotating side vortices, which are consistent with prior observations for *C. reinhardtii* (Drescher et al., 2010; Guasto et al., 2010). However, it failed to sustain the similarity throughout the beat cycle, particularly during the late power and recovery phases. This is because the flagellum is a slender filament, which adaptively changes its conformation in the later phase of the power stroke and the recovery stroke, likely to maximize propulsion and reduce drag. In contrast, in the model, the flagella sphere remains the same and experiences higher drag during recovery due to hydrodynamic interactions with the body sphere. This underscores the model's limitations in capturing phase-dependent asymmetries in force generation and drag. Thus, the standard model fails to reproduce accurate flow fields post the peak of the power stroke. Consequently, the time-averaged flow field does not compare well with experimental results, missing the front stagnation point and not having the vortices rearwards. Because in Stokes flow the instantaneous input power is converted entirely into viscous dissipation, a recovery phase with high dissipation indicates that more of the actuation is wasted overcoming drag/near-field coupling. The strong coupling thus increases the magnitude of backward motion, which cancels forward swimming. This is consistent with the defective mean swimming speed of the standard model, which is much lower than the reported swimming speed from experiments. The motivation for this work is thus to modify the model so that it can better represent both the swimming kinematics and the flow field.

The results from elliptical orbit modifications (figure 4) reveal that orbit geometry primarily affects the structure of the near-field vortex cores, with negligible influence on the far-field flow topology or swimming performance. Therefore, orbit modification alone appears insufficient to reconcile model predictions with experimental flow fields, indicating that further refinements to the swimmer model are necessary. Swimming kinematics analysis also suggests that circular orbit offers the highest swimming performance. This advantage arises from the hydrodynamic interaction between the spheres mediated by the surrounding flow. For circular orbit, the difference in the average distances from the flagellar spheres to the body sphere is the greatest between the power and recovery strokes. The differential resistance is thus maximized, leading to enhanced net forward propulsion. On the other hand, the orbit orientation does not alter the performance significantly.

The results from the introduction of phase-dependent torque profiles suggest that although phase-dependent torques may better approximate the biological asymmetry in force generation, they do not improve swimming performance within this model. This is because, when integrated over a full beat cycle, the phase-dependent torque averages to the same net input as the constant torque, leading to similar time-averaged velocity. The variable torque primarily redistributes work within the beat cycle rather than changing the total work per cycle, so its impact on the time-averaged flow and swimming speed is limited. Using a larger torque in the power stroke will increase the flow velocity proportionally. However, the duration of the power stroke will correspondingly decrease and that offsets the increasing flow strength. This trade-off results in the same dominance of the power stroke with respect to the recovery stroke in the cycle-averaged flow field. Thus, in this model framework, phase-dependent driving torque alone is not an effective mechanism for improving flow field agreement with experiment.

The most impactful modification is the introduction of a phase-dependent flagellar sphere size. By reducing the effective radius and thus drag during the recovery phase, we reproduce both the front stagnation point and the experimentally observed rearward side vortices (Drescher et al., 2010) (figure 8). This modification also monotonically increases swimming speed and efficiency as the size ratio decreases, indicating that the swimmer produces weaker backward motion during the recovery strokes and does not have to overcome larger drag. Moreover, having closer stagnation points and more rear-tilted side vortices as the size ratio decreases suggests that the instantaneous flow (movie 2) in the power stroke becomes increasingly dominant over that of the recovery stroke (see figure S2) and this dominance is unambiguously reflected in the time-averaged flow (figure 8). These results highlight that larger asymmetry (i.e., smaller recovery-phase sphere size) not only improves agreement with experimental flow fields but also enhances swimming performance.

An elliptical orbit introduces unequal curvature and nonuniform bead–body distances over a cycle, while a nonzero orbit rotation $\alpha$ redistributes forcing relative to the swimming axis. Although tuning differential drag $\lambda$ alone with standard circular orbit is sufficient to obtain desired flow features as shown in figure 6, the modification of orbit shape allows these features to be reproduced with less significant differential drag. Specifically, we can achieve closer stagnation points and tilted vortices with larger $\lambda$. Since the goal is to develop a minimal model, merely adding sphere size modification is logical. However, to match the location and orientation of side vortices with experiments, we must also change the scaffold geometry and orbit shape. Combining modified scaffold geometry, rotated elliptical orbit, and flagellar sphere size variation produces a side-oriented vortex pair and stagnation point position as observed experimentally (figure 8). The flow of the three-sphere model also agrees well with that of the classical three-Stokeslets model (Drescher et al., 2010). The optimized model not only fully captures the flow field but also ensures that the recovery stroke dissipates less energy than the power stroke (figure S4). Thus, the deviation from the experimental swimming velocity further reduces.

The fundamental influence on the flow field from the differential drag of flagellar spheres suggests that swimmer–swimmer hydrodynamic interactions will be altered. Consistent with this expectation, preliminary two-swimmer simulations show that, unlike the standard model where separation increases monotonically for the tested initial offsets, the modified model with only flagellar sphere recovery-to-power size ratio demonstrates that the distances between the two swimmers approach an equilibrium value (figure S5). This behavior is critically related to the front stagnation point at which the algal flow along the swimming axis switches direction. When the swimmers are initially separated beyond the stagnation point, the inward flow induced by the trailing swimmer resists the forward motion of the leading swimmer, resulting in a decrease in separation distance. Oppositely, when the swimmers are initially too close, the strong outward flow effectively repels them. An equilibrium separation is thus achieved. A systematic analysis of pair dynamics (e.g., dependence on phase offset, lateral displacement, and stability of the equilibrium distance) will be explored in future work.

In conclusion, our study suggests that matching the experimental flow field of *C. reinhardtii* within the three-sphere framework requires incorporating time-dependent drag asymmetries. This finding aligns with biological evidence that *C. reinhardtii* flagella undergo substantial changes in orientation and cross-sectional exposure to the fluid over the beat cycle, resulting in anisotropic drag forces that are essential for efficient swimming and realistic flow structure reproduction. Thus, our simple extension of the three-sphere model substantially improves the representation of algal flow fields and provides a valuable framework for studying alga–alga and alga–boundary interactions.

**Acknowledgements**

We thank Dr. Pragya Kushwaha for providing critical feedback on this study. We also thank Jonathan Blisko for contributions to preliminary studies related to this work. Portions of the manuscript text were drafted and/or edited with the assistance of ChatGPT (OpenAI, GPT-5.2 Thinking; accessed on [January–February 2026]). The tool was used for language polishing, rephrasing, and structural suggestions. The authors take full responsibility for the content, ensured originality, verified all technical statements, and confirmed that all cited literature is accurate and appropriately referenced.

**Funding statement**

This work was supported by the U.S. National Science Foundation (NSF) Biomechanics and Mechanobiology program (Awards 2242095 and 2242096/2438345).

**Competing interests**

The authors report no conflict of interest.

**Data availability**

The data and code that support the findings of this study are available from the corresponding author, X.Y., upon reasonable request.

**Author contributions**

M.I.Y.E. and X.Y. developed the theory. M.I.Y.E. performed simulations and analysis. G.R.P. contributed experimental data/interpretation. X.C. and X.Y. supervised research. All authors discussed the results. M.I.Y.E., X.C., and X.Y. wrote the manuscript.

**Supplementary Materials for**

**Improving Hydrodynamic Modeling of Free-Swimming Algae Using a Modified Three-Sphere Approach**

Md Iftekhar Yousuf Emon[1], Gregorius R. Pradipta[2], Xiang Cheng[2,3], Xin Yong[1]*

[1]Department of Mechanical and Aerospace Engineering, University at Buffalo, Buffalo, NY 14260, USA

[2]Department of Chemical Engineering and Materials Science, University of Minnesota, Minneapolis, MN 55455, USA

[3]Saint Anthony Falls Laboratory, University of Minnesota, Minneapolis, MN 55414, USA

*E-mail: xinyong@buffalo.edu

**Supplementary Figures**

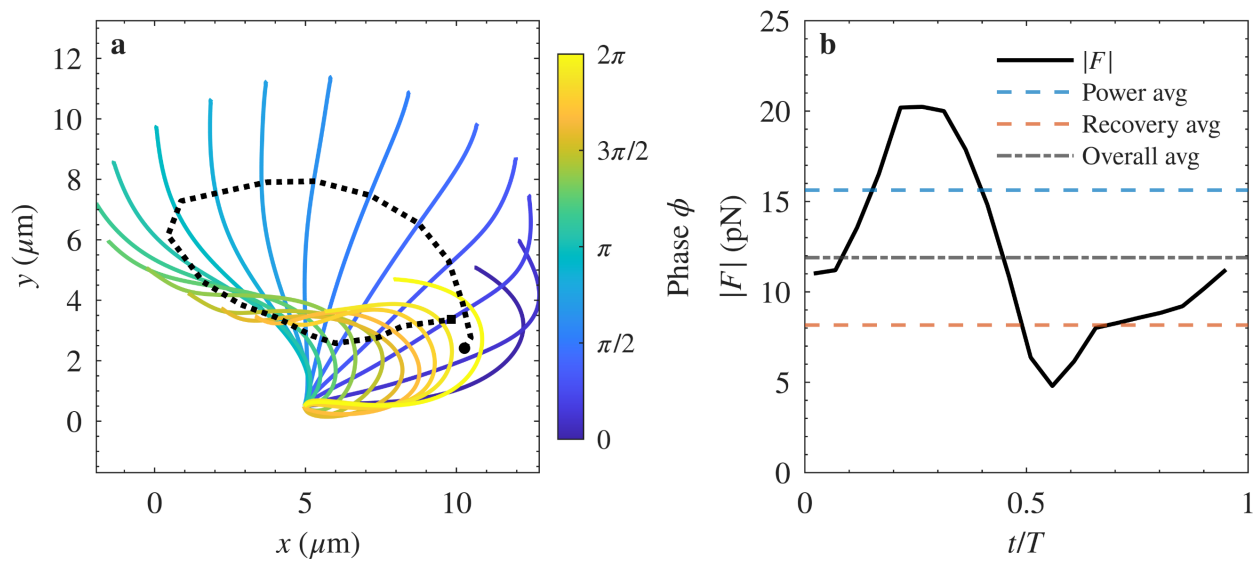

Figure S1: (a) Experimentally inferred flagellar beat pattern over one beat cycle. Colored curves show the instantaneous flagellar shapes at different phase angles $\phi \in [0,2\pi]$. The dotted curve denotes the trajectory of the flagellar centroid over a beat cycle. (b) Total flgellar force magnitude $|F|$ calculated based on the shape of the flagellum from the experimental data using the resistive force theory, showing larger forces during the power stroke than during the recovery stroke. Dashed lines denote the mean force over the power stroke (blue) and recovery stroke (red). The grey dash-dotted line marks the mean force over the full cycle.

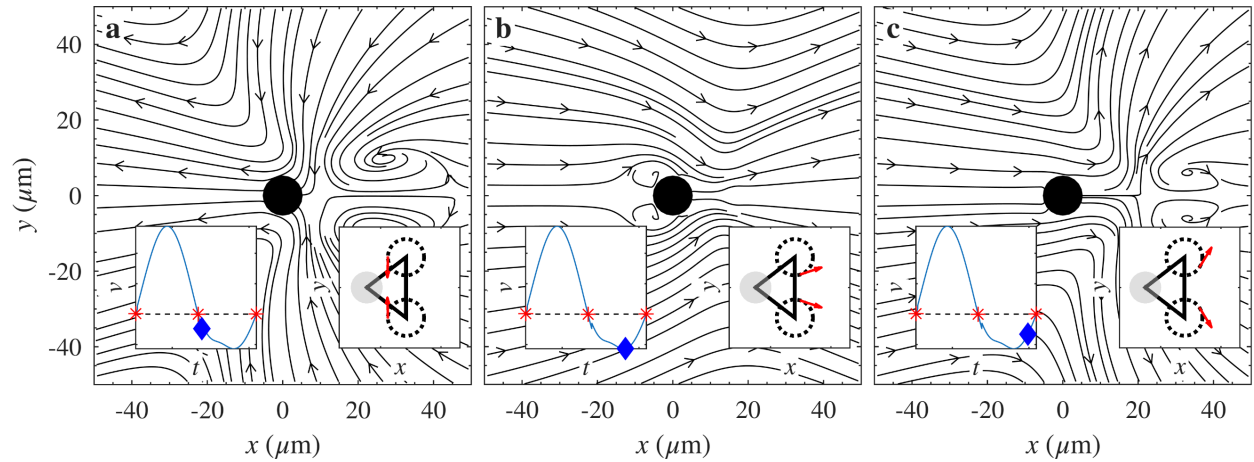

Figure S2: Time-resolved flow fields for the reduced size of flagellar sphere (size ratio 0.5) during recovery stroke. (a) At the onset of the recovery stroke, the driving torque is oriented nearly orthogonal to the swimming axis, producing distinct front vortices. (b) At the peak of the recovery stroke, the vortices shift rearward and become markedly weaker and confined to the surface of body sphere. (c) At the end of the recovery stroke, weak front vortices re-emerge, completing the cycle.

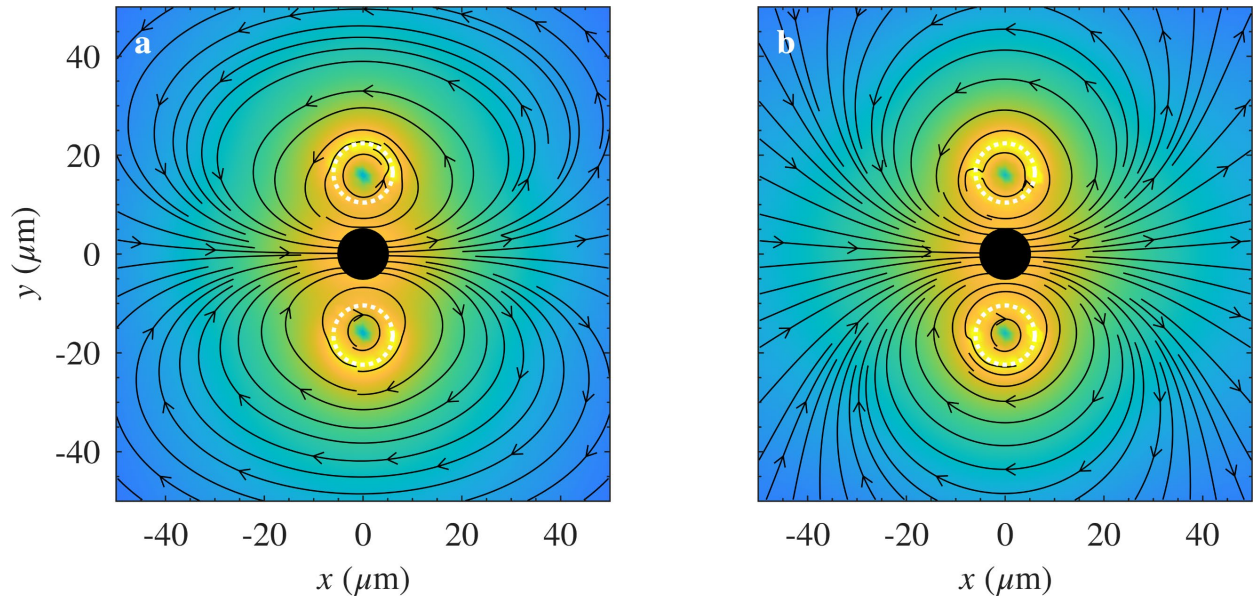

Figure S3: Cycle-averaged flow fields for the modified scaffold geometry ($h = 0$) with asymmetric drag of (a) $\lambda = 0.2$ and (b) $\lambda = 0.4$.

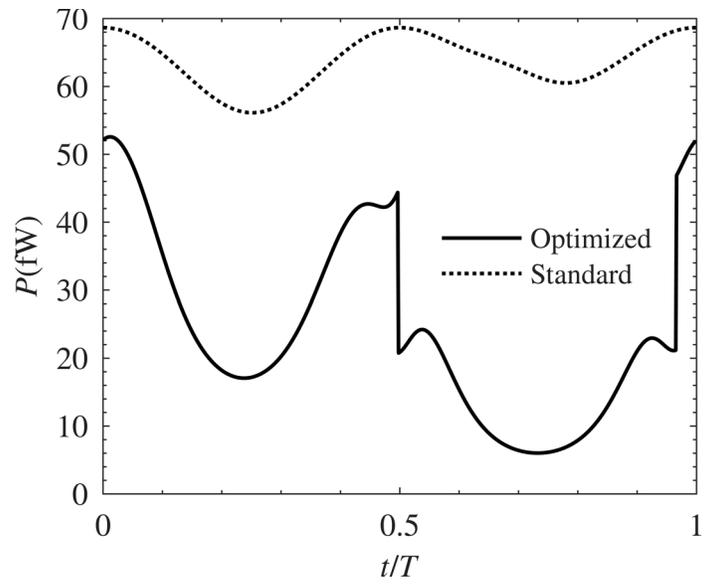

Figure S4: Viscous dissipation over one beat cycle for the standard and optimized model.

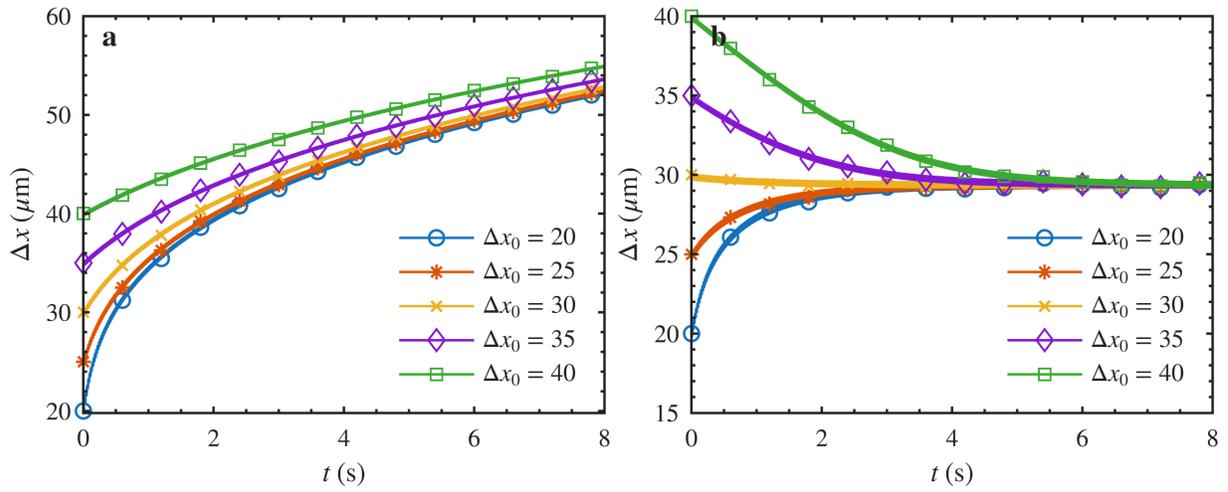

Figure S5: Time evolution of the separation distance between two interacting swimmers for different initial linear offsets for (a) standard three-sphere model and (b) modified three-sphere model with $\lambda = 0.5$.

**Supplementary Movies**

Movie 1: Time-resolved flow field over one complete flagellar beat cycle generated from the three-sphere swimmer model. Streamlines (black) depict the in-plane velocity directions in a swimmer-centered frame, while the background colormap shows the logarithmically scaled speed magnitude | **v** | (μm/s).

Movie 2: Time-resolved flow fields over one complete flagellar beat cycle generated from the optimized three-sphere swimmer model. The flagellar sphere radius varies with the phase to capture stroke-dependent drag asymmetry. The orbit shape, orientation, and position are also modified to adjust the position and angle of the side vortices. Streamlines (black) depict the in-plane velocity directions in a swimmer-centered frame, while the background colormap shows the logarithmically scaled speed magnitude | **v** | (μm/s).